\documentclass{epl}

\title{Electron-phonon coupling in the two phonon mode ternary alloy $Al_{0.25}In_{0.75}As/Ga_{0.25}In_{0.75}As$ quantum well}
\author{C. Faugeras\inst{1} \and G. Martinez\inst{1} \and F. Capotondi\inst{2,3} \and G. Biasiol\inst{2} \and L. Sorba\inst{2,3}}
\institute{
  \inst{1} Grenoble High Magnetic Field Laboratory, Max Planck Institut f\"{u}r Festk%
\"{o}rperforschung\\
and Centre National de la Recherche Scientifique - BP 166, 38042
Grenoble Cedex 9, France.\\
  \inst{2} NEST-INFM and Laboratorio Nazionale TASC-INFM - Area
Science Park, I-34012 Trieste, Italy.\\
\inst{3}  Dipartimento di Fisica, Universita di Modena e Reggio
Emilia, I-41100 Modena, Italy.}

\begin{document}

\maketitle

\begin{abstract}
We have investigated the infrared transmission of a
two-dimensional (2DEG) electron gas confined in a
$Al_{0.25}In_{0.75}As/Ga_{0.25}In_{0.75}As$ single quantum well in
order to study the electron optical phonon interaction in a two
phonon mode system. Infrared transmission experiments have been
performed in both the perpendicular Faraday (PF) and tilted
Faraday (TF) configurations for which the growth axis of the
sample is tilted with respect to the incident light propagation
direction and to the magnetic field direction. The experimental results
lead to question the validity of the concept of polaron mass in a
real material.
\end{abstract}

\section{Introduction}
Based on the Fr\"{o}hlich interaction mechanism, polaronic effects
have attracted much interest both on the theoretical and on the
experimental side \cite{Devreese}. In polar semiconductors, the
propagation of longitudinal optical (LO) phonons produces a
macroscopic electric field which couples -via the Fr\"{o}hlich
interaction- to the electron motion. This interaction is predicted
to give rise to the Resonant Magneto Polaron (RMP) effect between
the $|n=1>$ Landau level and the $|n=0 + 1 LO>$ state, n being the
Landau level index. This RMP effect, in polar materials, is
predicted to induce an anti-crossing of the cyclotron resonance
(CR) transition when the CR frequency $\omega_c=eB/m^*$, where $B$
is the magnetic field and $m^*$ is the electron effective mass, is
tuned through the LO phonon frequency \cite{Dassarma,Larsen}. It
has however never been directly observed and different studies on
a $Ga_{0.47}In_{0.53}As$ heterostructure \cite{Nicholas85}, and
more recently, on a GaAs quantum well \cite{Poulter01} and on a
InAs quantum well \cite{Hu03}, indicate that no interaction
affects the CR mode in the vicinity of the LO phonon energy. It
has recently been shown, based on infrared transmission
measurements performed on single GaAs quantum wells (QW) doped
with carrier concentrations ranging from $6\times10^{11}$ to
$12\times10^{11}$ cm$^{-2}$ \cite{Faugeras04}, that the CR mode
does not couple to the LO phonon of the quantum well but instead,
couples to the hybrid inter electric sub-band plasmon - LO phonon
mode that develops in a doped quasi two dimensional layer. These
results have been explained using a semi classical model based on
the dielectric function formalism \cite{Bychkov}. This model
reproduced the experimental data in an accurate way without
assuming any specific electron-phonon interaction. The observed
coupling is then of dielectric origin. In these GaAs QW studies,
the inter electric sub-band energy was approximately twice larger
than the LO phonon energy. It is then of fundamental importance to
know whether these findings are specific of the GaAs QW or are
indeed more general. We report here on results obtained with a
$Al_{0.25}In_{0.75}As/Ga_{0.25}In_{0.75}As$ QW sample differing
from the GaAs QW by two main characteristics: the phonon structure
is a two mode system and the inter sub-band energy is close to the
GaAs-like LO phonon energy. After describing the sample and the
experimental details, we present the experimental results, their
interpretation and discuss the results.

\section{Samples and experimental details}
\begin{sloppypar}
The sample $HM617$ is a single $30$ nm wide $Ga_{0.25}In_{0.75}As$
quantum well grown by molecular beam epitaxy on a (001) GaAs
substrate and sandwiched between two $Al_{0.25}In_{0.75}As$
barriers. To achieve a good electron mobility, a step graded
buffer $Al_{1-x}In_{x}As$ with x ranging from $0.15$ to $0.85$ was
grown on the substrate to obtain a quasi complete strain
relaxation in the quantum well. The sample is unintentionally
doped \cite{Capotondi04} and the electron density $n_{s}$
determined from low temperature transport measurements is of
$2.8\times10^{11} cm^{-2}$ with a mobility of $20$
$m^{2}.V^{-1}.s^{-1}$. $Ga_{0.25}In_{0.75}As$ is a two-phonon mode
system with two distinct phonon types, InAs-like phonons with
energy $\hbar\omega_{TO}=21.7$ meV and $\hbar\omega_{LO}=22.4$
meV, and GaAs-like phonons with energy $\hbar\omega_{TO}=27.9$ meV
and $\hbar\omega_{LO}=29.5$ meV. Preliminary calculations indicate
that the inter electric sub-band energy, for this particular
sample, should be of the order of the GaAs-like LO phonon energy,
a situation which has never been explored yet.
\end{sloppypar}
The experiments are performed with a Fourier transform Brucker
IFS-113 spectrometer and the signal is detected by a silicon
bolometer located behind the sample at a temperature of $1.8$ K. To
avoid interference effects, the sample has been wedged with an
angle of $2 ^o$. Magneto infrared experiments have been performed
in an absolute way by using a rotating sample holder which allows
 to switch  \textit{in situ} between the sample and a reference (in
the present case a GaAs substrate). We can then determine the
absolute transmission of the sample which is the ratio of the
transmission of the sample at a given field by the transmission of
the reference at the same value of the magnetic field. Magnetic
field dependant absorptions, of electronic origin, are more
visible on the relative transmission spectrum which is the ratio
of the absolute transmission at a given magnetic field by the
absolute transmission at zero field. In the tilted Faraday (TF)
configuration, for which the normal to the sample surface is
tilted by an angle $\theta$ with respect to the magnetic field and
to the light propagation direction, the angle is imposed
mechanically by a wedged sample holder. The CR absorption is
measured using a superconducting magnet at magnetic fields up to
$13$ T. The interpretation of the experimental data is
accomplished by comparing them with the results of a
multi-dielectric simulation of the entire sample structure by an
appropriate model \cite{Bychkov}. This is essential because in the
optical phonon range of energy, spectra can be strongly distorted
by interference effects independent of any electron-phonon
interaction. Probably the most studied III-V material is the GaAs
quantum well for which transmission results are always obscured by
the strong absorption of the substrate. The $Ga_{0.25}In_{0.75}As$
quantum well system is more attractive to study polaronic effects
because the optical phonon energies of this compound are below the
GaAs  \textit{reststrahlen} band : this allows to study the
infrared absorption through the interesting optical phonon range
of energy.

\section{Experimental results}

\begin{figure}

 \twofigures[scale=0.45]{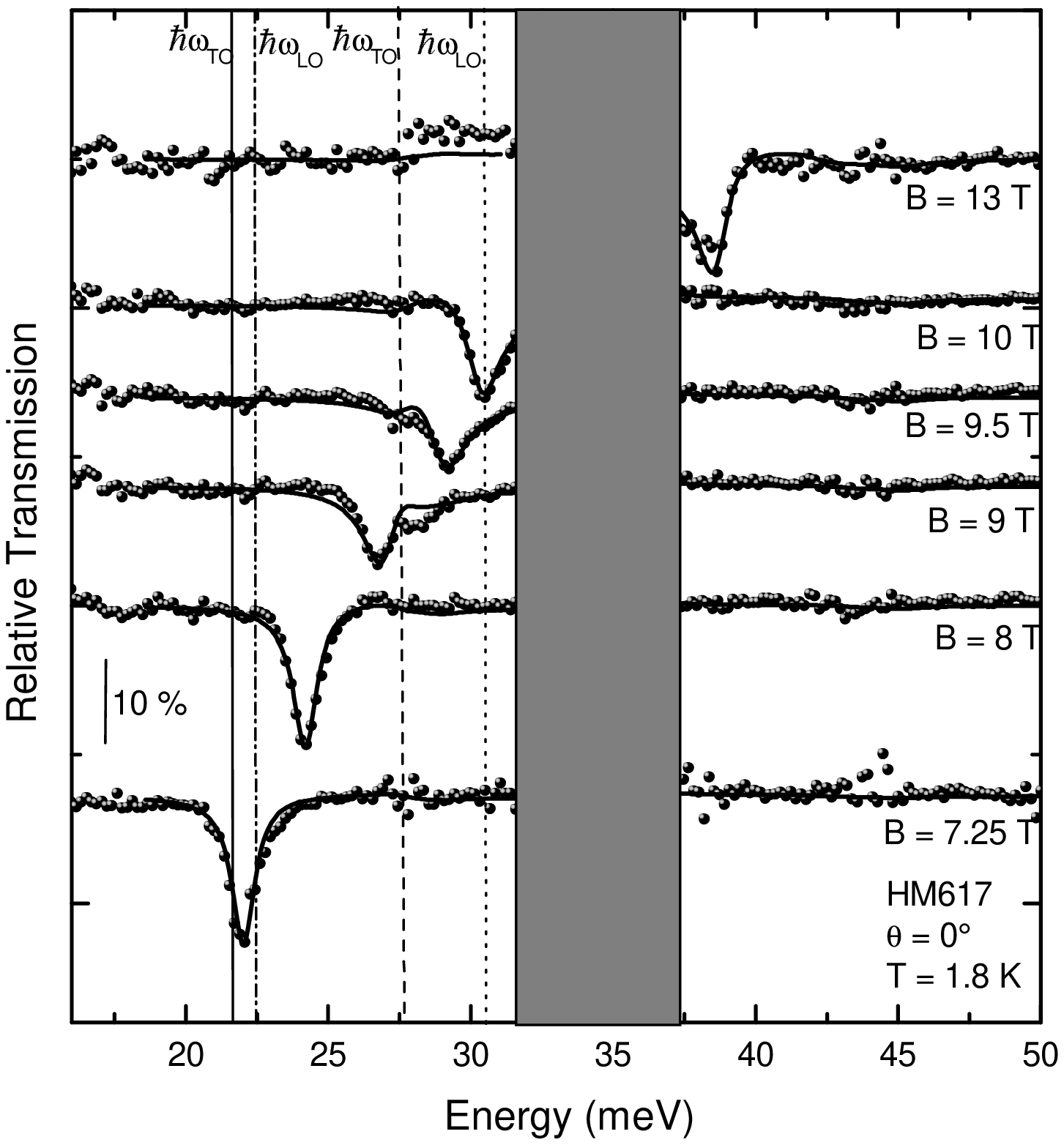}{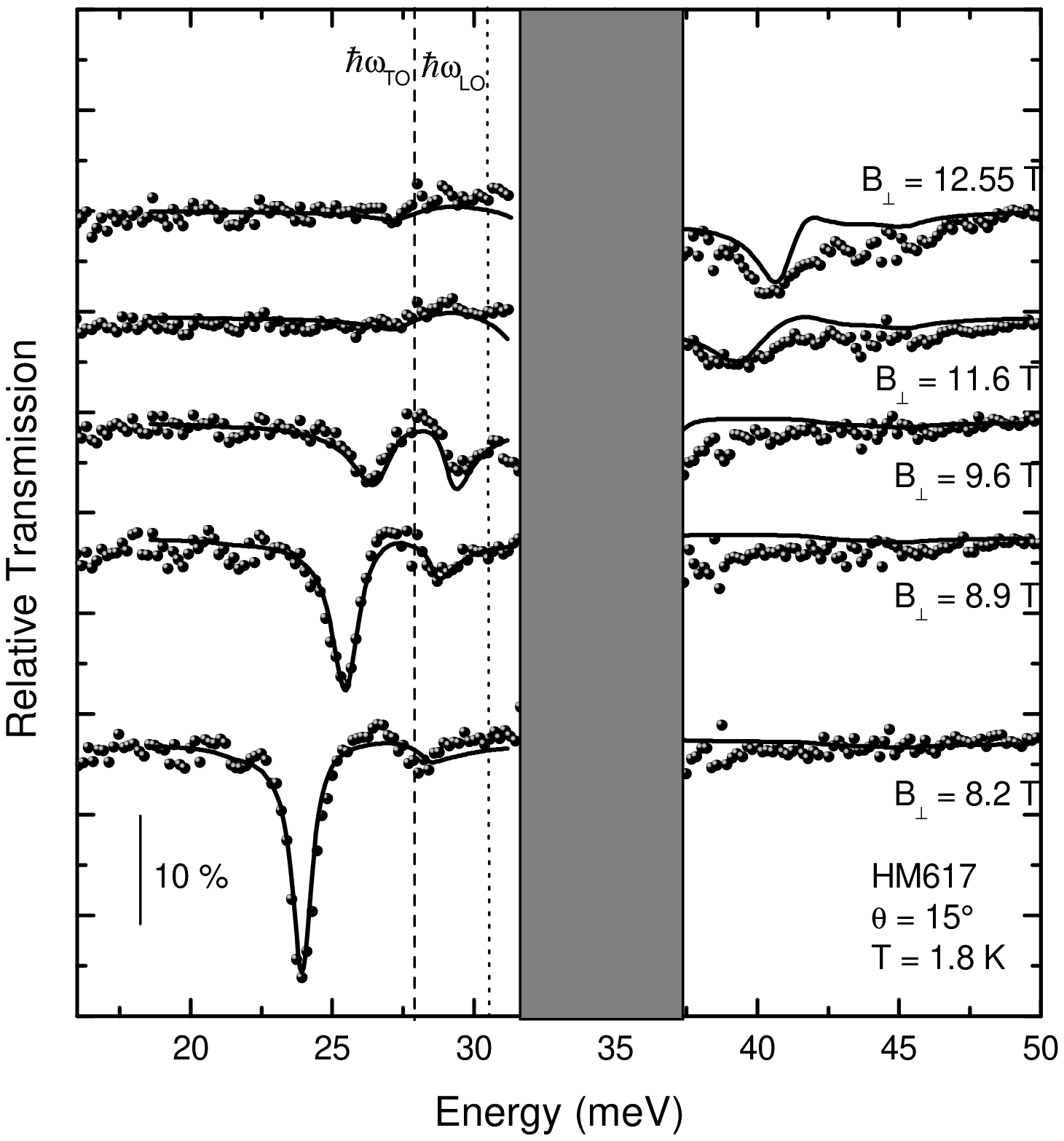}

  \caption{ Experimental CR spectra
  (black dots) and simulated spectra (solid lines) of
sample HM617 measured at $T=1.8$ K in PF configuration. The
InAs-like TO phonon is indicated by the vertical solid line, the
InAs-like LO phonon by the vertical dash-dotted line, the
GaAs-like TO phonon the vertical dashed line and the GaAs-like LO
phonon by the vertical dotted line. The shaded region of energy
corresponds to the  \textit{reststrahlen} band of the GaAs substrate.}
  \label{fig1}
  \caption{ Experimental CR spectra (black dots) and simulated spectra (solid lines) of
sample HM617 measured at $T=1.8$ K in TF configuration for
$\theta=15^o$. }
 \label{fig2}

\end{figure}

In the perpendicular faraday (PF) configuration, the spectra
exhibit a characteristic field dependent CR absorption line (see
black dots in fig.~\ref{fig1}). This CR absorption is tuned
through the entire range of InAs-like and GaAs-like optical phonon
energies without being affected by neither the InAs-like LO phonon
(dash-dotted line in fig.~\ref{fig1}) nor the GaAs-like LO phonon
(dotted line in fig.~\ref{fig1}) of the quantum well. Therefore,
in this PF configuration, the expected polaron coupling is not
observed. This is in accordance with results obtained on GaAs QW
\cite{Faugeras04}. At low fields below $\approx 10$ T, the
simulation (solid lines in fig.~\ref{fig1} and fig.~\ref{fig2})
provides a fitted value of the effective mass $m^*=0.038$ $m_0$ in
agreement with previous electrical detection of CR experiments
performed on the same type of samples \cite{Desrat04}. When the
magnetic field is tuned to $13$ T, the CR absorption is observed
above the \textit{reststrahlen} band of the GaAs substrate with a
significant increase of the mass $m^*$ which, in the present case,
reaches a value of $0.0395$ $m_0$. This may be due to band non
parabolicity effects known to be significant for this compound.

The infrared transmission measured in the TF configuration shows a
very different behavior (see black dots in fig.~\ref{fig2}): When
the CR energy is well below the GaAs-like optical phonon range of
energy, we observe a single electronic absorption that pins at an
energy of $27.2\pm0.3$ meV while a second absorption appears at an
energy of $28.3\pm0.4$ meV. This second absorption then enters the
obscured region of the GaAs substrate
 \textit{reststrahlen} band while a third absorption appears above
this band. For a given perpendicular component of the magnetic
field, the minima of transmission, measured in the TF
configuration above the  \textit{reststrahlen} band of the
substrate, is well above the one measured in the PF configuration
(see spectra at $13$ T in fig~\ref{fig1} and at $12.55$ T in
fig~\ref{fig2} and full dots in fig.~\ref{fig3}) which is
surprising. Indeed the cyclotron energy is usually believed to
depend only on the perpendicular component of the magnetic field
which should lead to a \textbf{lower} energy of the CR than the
one measured in the PF configuration as this is observed in GaAs
QW \cite{Faugeras04}. Moreover, the width of this second
absorption is much bigger than the one of the CR mode measured in
the PF configuration in the same range of energies.

We argue that all the observed effects are due to the coupling of
the CR mode with the hybrid inter sub-band plasmon - LO phonon
modes as observed in a GaAs single quantum well \cite{Faugeras04}.
In the present study however, the polar slab containing the 2DEG
has two distinct phonon modes (InAs-like and GaAs-like) which are
both coupled to the plasmon inter sub-band with the inter sub-band
energy close to that of the GaAs like LO energy.

\section{Interpretation of the results}

To explain these data, we have calculated the optical transmission
of the whole structure using the multi-dielectric simulation
\cite{Bychkov} and the exact layer sequence for our sample. The
calculation consists in evaluating for each layer N with a
dielectric tensor $\varepsilon_{N}(\omega)$, the transfer matrix
$M_{N}$. This requires to solve the Maxwell equations in order to
determine the appropriate modes of propagation of light
$\overrightarrow{k}^{i}_{N}$  inside the layer. There are four
modes $\overrightarrow{k}^{i}_{N}$ for each layers. Writing the
conservation relations of the tangential components of both
electric and magnetic fields at each interface enables us to
determine the $4\times4$ transfer matrix characterizing a layer.
If this layer does not contain free charges,
$\varepsilon_{N}(\omega)$ is then diagonal and the $4\times4$
transfer matrix decomposes into two blocks of $2\times2$ matrices
related to the transverse electric (TE) and transverse magnetic
(TM) modes respectively. For a doped layer in TF configuration,
all the components of $\varepsilon_{N}(\omega)$ and therefore of
the transfer matrix are non zero and the TE and TM modes are
mixed. The transmission of the whole structure can then be
calculated as soon as the dielectric tensor of the 2DEG in the TF
configuration has been evaluated.

The main effect of the in-plane component of the magnetic field is
to couple the x- and z- part of the wave function describing the
electron gas. This coupling can be evaluated using perturbation
theory in the case of a square quantum well. The use of a
parabolic confinement \cite{Maan84, Merlin87}, characterized by an
inter sub-band energy $\hbar\Omega$, provides an exact solution of
this problem. It is this model, based on the Drude formalism, that
we have used to evaluate, neglecting retardation effects, all
components of the dielectric tensor $\varepsilon_{QW}(\omega)$
describing the quantum well and then to simulate the transmission.

The results of such a calculation for the PF configuration are
presented in fig.~\ref{fig1} (solid lines). When the cyclotron
energy is tuned through the GaAs-like optical phonon range of
energy, the electronic absorption splits around the GaAs-like TO
phonon frequency (see spectrum at $9$ T in fig.~\ref{fig1}). This
behavior was first attributed to a dielectric artifact
\cite{Ziesmann87,Karrai88} as the TO phonon is a natural pole of
the dielectric function describing the 2DEG, but this effect seems
to be more intrinsic. As the dielectric calculation, in which no
electron-TO phonon interaction is considered, cannot reproduce the
experimental spectra, the observed discontinuity of the cyclotron
energy at the TO energy is a sign of an interaction of the CR mode
with the GaAs-like TO phonon mode of the $Ga_{0.25}In_{0.75}As$
quantum well. The same effects were already observed in this kind
of compounds \cite{Nicholas85} and in GaAs QW \cite{Faugeras04}.
 Therefore it seems that in doped QWs, there exists in general an
 interaction between the cyclotron mode and the TO phonon modes.
 This deserves to be worked out on theoretical grounds. The average
 electron scattering time deduced from the fitting procedure is
 $\tau_{CR}=6\times10^{-12}$ s which corresponds to a mobility of
 $27$ $m^{2}.V^{-1}.s^{-1}$ in agreement with the mobility deduced from
transport measurements.

\begin{figure}

\includegraphics[scale=0.90]{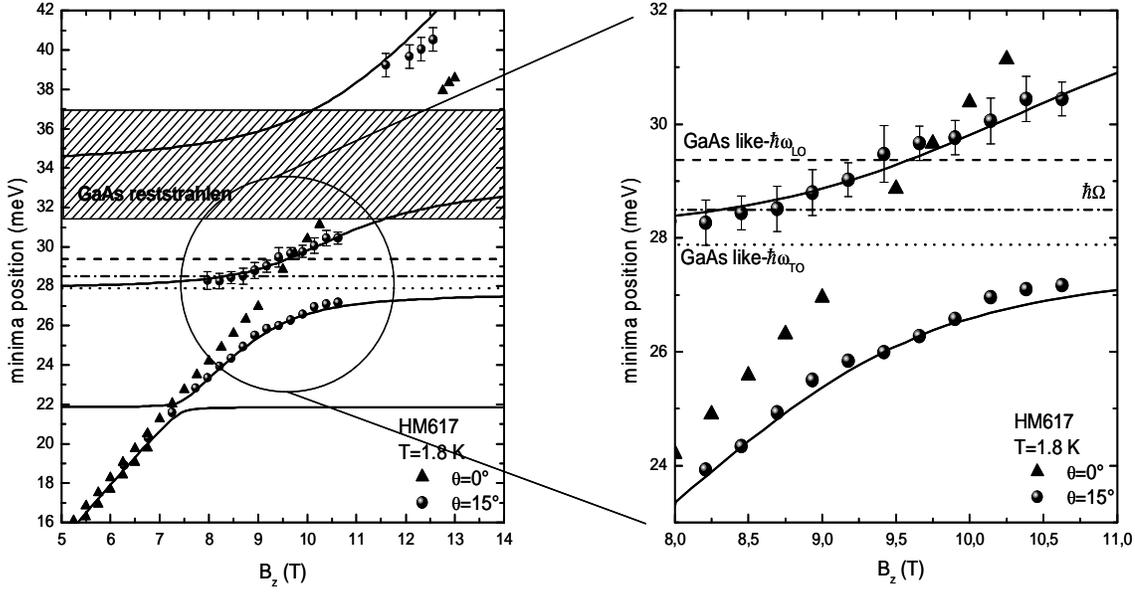}
\caption{ Evolution of the energies of the zeros of
$\epsilon_{zz}(\omega)$ (solid lines) and of the minima of
transmission measured in TF configuration with $\theta=15^o$
(black dots) and in PF configuration (black triangles) as a
function of the perpendicular component of the magnetic field. The
GaAs-like LO phonon energy (dashed line), the GaAs-like TO phonon
energy (dotted line) and the quantum well inter sub-band energy
(dash-dot line) are indicated. The hashed region corresponds to
the GaAs substrate  \textit{reststrahlen} band.} \label{fig3}
\end{figure}

The use of the TF configuration enables to excite the 2DEG with a
non vanishing component of the incident electric field
perpendicular to the quantum well. In this case, the response of
the electronic system involves the $\epsilon_{\alpha z}(\omega)$
($\alpha=x,y,z$) component of the dielectric tensor.
 For instance, the $\epsilon_{zz}(\omega)$ component describing the 2DEG
 confined in a two phonon mode polar slab can be formulated, ignoring for
 simplicity the imaginary part, as
\cite{Zheng02,Bychkov}:
\begin{equation}
\epsilon_{zz}(\omega)=\epsilon_0\prod_i\frac{\omega_{Li}^2-\omega^2}{\omega_{Ti}^2-\omega^2}+
\frac{\omega_p^2(\omega_{cz}^2-\omega^2)}{\omega^4-\omega^2(\Omega^2+\omega_c^2)+\omega_{cz}^2\Omega^2}
\end{equation}
where $\epsilon_0$ is an average static dielectric constant taken
to be equal to $13.85$, L and T refer to the longitudinal and
transverse phonons while the subscript i refers to the relevant
GaAs-like and InAs-like modes, $\omega_p$ is the plasma frequency,
$\Omega$, the inter sub-band frequency, $\omega_c$,  the cyclotron
frequency and $\omega_{cz}=eB_{\perp}/m^*$, with $B_{\perp}$ being
the perpendicular component of the magnetic field. This component
of the dielectric tensor has $4$ zeros corresponding to
longitudinal solutions: for low values of the magnetic field, they
correspond to the CR like mode and to three hybrid inter electric
sub-band plasmon-LO phonon modes of the quantum well. For higher
values of the magnetic field, the evolution of these modes is far
from trivial: they are displayed, as full lines, in
fig.~\ref{fig3}, as a function of $B_{\perp}$ for a tilt angle of
$15^o$. One can compare these results with the observed minima of
transmission (full dots in fig.~\ref{fig3}). Though all non
diagonal elements of the dielectric tensor play also a role in the
calculation of the transmission, the main contribution is provided
by $\epsilon_{zz}(\omega)$. Therefore the agreement between the
experimental results and the solutions of Eq. 1 is very good.

In the TF configuration, we impose, to fit the experimental data,
that the effective mass for a given perpendicular component of the
magnetic field is the same as the one deduced in the PF
configuration for the same value of the magnetic field. We then
fit the inter sub-band energy $\hbar\Omega$ together with the
inter sub-band scattering time ($\tau_{IS}$). The deduced inter
sub-band energy is $\hbar\Omega=28.5$ meV. The fitted spectra are
obtained with a value of $\tau_{IS}=3\times10^{-12}$ s, much
shorter than $\tau_{CR}$ resulting in a larger width for the
corresponding transmission structure. This line width dominates
the absorption in the GaAs-like optical phonon energy range. This
situation is different from the previously measured absorption in
the GaAs QW \cite{Faugeras04} where the coupling involved the
GaAs-like phonon hybrid mode with a width considerably less than
that of the inter sub-band plasmon observed in this report. The
reason is that the hybrid mode which couples to the CR transition
has a phonon character when $\hbar\Omega\gg\hbar\omega_{LO}$
whereas it acquires an inter sub-band transition character when
$\hbar\Omega\preceq\hbar\omega_{LO}$ as in the present case. This
inter sub-band transition is known to appear with a much larger
line width \cite{Liu} than that observed for the resonance
transition between non dispersive levels as Landau levels or
phonons.

Despite small discrepancies, the multi-dielectric simulation in
the TF configuration reproduce the main observed features, which
are the anti-crossing behavior below the GaAs-like LO phonon
energy with the pinning of the CR absorption, and the appearance
of a third absorption at an energy higher than the observed
cyclotron energy in PF configuration. This means that, though
simple, our model contains all the physics necessary to explain
the observed behavior without invoking any electron-LO phonon
interaction.

Therefore, the good agreement between the theory and the
experiment allows to assign unambiguously the origin of the
observed anti-crossing in the TF configuration. It is clearly due
to the dielectric coupling of the CR with each of the inter
sub-band plasmon-LO phonon hybrid modes. Though, due to the small
oscillator strength of the InAs-like mode and the relatively large
width of the cyclotron line, this coupling is not observed for the
lower hybrid mode around $22$ meV, it is clearly visible around
the second hybrid mode ($27.8$ meV). Therefore, from the present
results, and those previously obtained in GaAs QW
\cite{Faugeras04}, one can say that, in 2D doped QW (at least in
III-V compounds), whatever is the phonon structure, the carrier
density or the relative value of the inter sub-band energy with
respect to the LO phonon energies, the experimental results can be
well reproduced with the proposed model.

\section{Discussion of the results}

As discussed in the preceding section, the model used to evaluate
the dielectric function neglects retardation effects: the
dielectric function is therefore formulated at $k \rightarrow 0$.
This k-dependence has been studied by Peeters \textit{et al.}
\cite{Peeters87} in this system and it appears to be significant
for k-vectors of the order of $10^{6}$ cm$^{-1}$, decreasing
rapidly below this value. In the present experiments, performed in
the infra-red region, the real part of the k-vector at the CR
frequency is of the order of $10^{4}$ cm$^{-1}$ and therefore
neglecting retardation effects in the derivation of the dielectric
response is well justified.

Since when deriving the multi-dielectric model of transmission
\cite{Bychkov}, one has to solve the Maxwell equations for each
layer, we are indeed using the polariton picture with a non
k-dispersive dielectric function. What is not included in the
model is any electron-phonon interaction which should appear in
the dielectric response function. In the case of the Fr\"{o}hlich
interaction, it could be included through the memory function
    approach \cite{Peeters85} which predicts an anti-crossing of the CR
transition around the LO energy in the PF configuration which
should be significant for the carrier concentration used here:
this RMP effect is however not observed. Since it is directly
related to the concept of polaron mass, this concept itself can be
questioned in real doped systems.

It is easy and instructive to extend the present model to the
3-dimensional case by letting $\hbar\Omega\rightarrow 0$ while
keeping $ \omega_{p}^{2}$ constant. As an example, in this limit,
the zeros of Eq. 1, for a one-phonon mode system at $B=0$ are
simply the well known $\omega^{\pm}$ modes found for the
plasmon-phonon coupling in a polar semiconductor, as first
reported by Mooradian \textit{et al.} \cite{Mooradian66}.

\section{Conclusion}
In conclusion, infrared transmission experiments have been
performed on a single $Al_{0.25}In_{0.75}As/Ga_{0.25}In_{0.75}As$
quantum well to investigate the nature of the electron-phonon
coupling in a two phonon modes system. In PF configuration, we
observe no sign of interaction of the CR mode with the GaAs-like
LO phonon. Instead, we observe signs of a coupling with the TO
phonon of the quantum well. As in pure GaAs single quantum well,
but changing the physical parameters of the problem, we observe
for the first time, the coupling of the CR mode with the two
hybrid GaAs-like LO phonon-inter electric sub-band modes of the
quantum well with free carriers. This effect is of pure dielectric
origin and is not due to any specific electron-phonon interaction.

\acknowledgments The GHMFL is "Laboratoire conventionn\'{e} \'{a}
l'UJF et l'INPG de Grenoble". The work presented here has been
supported in opart by the European Commission through the Grant
HPRI-CT-1999-00030.

\end{document}